\documentclass[aip,preprint]{revtex4-1}
\usepackage{graphicx}\usepackage{amsmath}\usepackage{amssymb}
\begin{document}

\title{A battery-operated, stabilized, high-energy pulsed electron gun for the production of rare gas excimers} 
\author{L. Barcellan}
\affiliation{INFN Section, Padua, Italy}
\author{E. Berto}
\affiliation{Department of Physics, University of Padua, Italy}
\author{G. Carugno}
\affiliation{INFN Section, Padua, Italy}
\author{G. Galet}
\affiliation{Department of Physics, University of Padua, Italy}
\author{G. Galeazzi}
\affiliation{Department of Physics, University of Padua, Italy}
\author{A. F. Borghesani}
\email[]{armandofrancesco.borghesani@unipd.it}
\affiliation{CNISM Unit, Department of Physics, University of Padua, Italy}
\affiliation{INFN Section, Padua, Italy}

\begin{abstract}
We report on the design of a new type of electron gun to be used for experiments of infrared emission spectroscopy  of rare gas excimers. 
It is based on a filament heated by means of a pack of rechargeable batteries floated atop the high-voltage power supply.
The filament current is controlled by a feedback circuit including a superluminescent diode decoupled from the high voltage by means of an optical fiber.
Our experiment requires that the charge injection is pulsed and constant and stable in time.
This electron gun can deliver several tens of nC per pulse of electrons of energy up to $100\,$keV into the sample cell. 
This new design eliminates ripples in the emission current and  ensures up to 12 hrs of stable performance.
\end{abstract}
\maketitle 

\section{Introduction}\label{Intro}
Collisions between high-energy electrons and atoms or molecules may lead to the creation of a rich number of excited and reactive species. These physical events are exploited in several experimental techniques and processes including pollutant controls \cite{Newson:1995uq,kohno98}, excitation for lasers \cite{rhodes1979}, enhancement of chemical reaction rates \cite{mason2009} and many more \cite{mason2005}. 

In particular,  in Xe-based ionizing radiation detectors for high-energy physics \cite{knoll}, Xe atoms are excited by collisions with secondary electrons released by the passage of an ionizing particle through the gas. The excited atoms may collide with neutral, ground-state Xe atoms leading to the formation of Xe$_{2}$ molecules in an excited state ({\it excimers})  \cite{koe1974}. These, then, spontaneously decay towards the dissociative ground state by releasing the 172 nm ultraviolet (UV) photon whose detection tags the passage of the ionizing particle.

In previous experiments,  we  discovered that the excitation of Xe atoms by impact with high-energy electrons and the subsequent collision between an excited- and one ground-state atom also lead to the production of Xe$_{2}$ excimers in molecular excited states higher in energy than the first excited one. These higher lying molecular states give origin to an infrared (IR) luminescence upon transitions to lower lying states in the energy degradation pathway that eventually leads to the dissociation of the Xe$_{2}$ molecule. The spectral investigation of such IR luminescence, also as a function of the gas density, has allowed the researchers to assign the molecular states involved in the transition and to elucidate the nature of the interaction of the excimer with its environment \cite{Borghesani2001,borghesani2007jpb}.

In our experiment, high-energy electrons were produced in a homemade electron gun in a diode configuration. Details can be found in literature \cite{Borghesani2001}. Here, we only recall that electrons were extracted out of a brass photocathode by a 15 ns-short pulse of an UV ArF excimer laser at 193 nm with a repetition rate up to 200 Hz. The electrons were accelerated into an evacuated tube by keeping the cathode at a potential in  the range up to 100 kV negative with respect to the grounded vacuum housing.

After a magnetic focusing step accomplished by a first magnetic coil, electrons were deflected by 90$^{\circ}$ with the aid of a second magnetic coil in order for them to enter into the reaction cell containing the high-pressure sample gas. The 90$^{\circ}$ bend is necessary not to block the laser light impinging on the photocathode, otherwise photoextraction would be inhibited. By so doing, even small misalignments in the long and curved path travelled by the electrons before entering the sample cell had the consequence that the fraction of the photoextracted electrons eventually injected into  the cell was quite small. We estimated that the charge finally injected into the gas was of the order of a few tens of pC per pulse or less. Moreover, the laser intensity, hence the charge extraction, decreased with time up to $5\%$ per hour.

This small amount of charge is sufficient to produce a detectable IR
 signal because of the large photon yield of Xe. However, it is mandatory to increase the amount of charge injected into the gas during each pulse in order to extend our investigation on the density dependence of the IR luminescence emitted by the decay of different rare gas excimers with a smaller photon yield than Xe.

Owing to this reason, we have decided to develop an electron gun of different design that can be operated in repetitive, single-shot mode.  
The main requirements that the new electron gun has to satisfy in order to comply with the needs of our experiment are
the following:
\begin{itemize}
\item [1] it has to be operated in repetitive, single-shot mode at a repetition rate $100<f<200\, $Hz;
\item [2]it must deliver several nC per pulse of electrons at about 70 kV;
\item [3] the delivered charge must be quite constant during several hours of operation;
\item [4] it must be quite compact and easy to operate.
\end{itemize}
We have, thus, devised an electron gun based on a dc powered hot-filament cathode in a simple diode configuration whose emission current can be controlled by means of a feedback circuit so as to ensure the required stability. In the following, we will described the design of this new electron gun.

\section{Description of the apparatus}

The main reason for the smallness of the charge injected into the sample cell using the previous electron gun is that the electrons photoextracted by the excimer laser  were accelerated in the direction of the laser beam itself. By so doing, the sample cell had to be located sideways with respect to the laser beam in order no to block it and electrons had to be magnetically deflected by 90$^{\circ}.$

The most obvious improvement to increase the amount of charge injected into the  cell is to avoid the 90$^{\circ}$ bend required by the old design. This requirement rules out the possibility of extracting electrons by means of the UV excimer laser unless light is guided onto the photocathode at an angle by means of an UV-grade optical fiber that leads to a reduced photoextraction.  In addition to that, we have also to mention the fact that the excimer laser requires a careful and heavy-duty maintenance in order to keep it in reliable working conditions and its energy output is not as stable in time as required but steadily decreases by a few \% per hour.

So, the simplest way to enhance the charge injection, and to reduce the troubles with the excimer laser, is to replace electron photoextraction with thermionic emission by a hot filament for which no bending of the beam is required and the sample cell can directly be located along the emission direction.

To this goal, we have designed an electron gun in a diode configuration similar to a scheme previously described in literature \cite{peterson1992}. 
In figure \ref{fig:blockdiagram} we show a simplified block diagram of the electron gun assembly.

\subsection{High voltage- and filament power supply}\label{sec:HV}
Electrons are released by thermionic emission from a  thoriated tungsten filament of the type used in electron microscopes which is directly heated by Joule effect.
The power supply for an electron gun of this kind typically consists of a dc HV power supply, an isolation transformer, and an ac filament power supply. This kind of setup is not very well suited to our purposes for several reasons. The first one is that isolation transformers are very bulky objects whereas we want to keep the electron gun at a minimum size as possible. Moreover, an ac filament power supply always introduces unwanted ripples into the electron beam current that might make more difficult, if not even impossible, the analysis of the IR signal by means of the FTIR interferometer. 
So, we have decided to use dc sources for both the HV- and the filament power supplies. This choice represent the main novelty of the present design. 

\begin{figure}[htbp]
 \includegraphics[width=\textwidth]{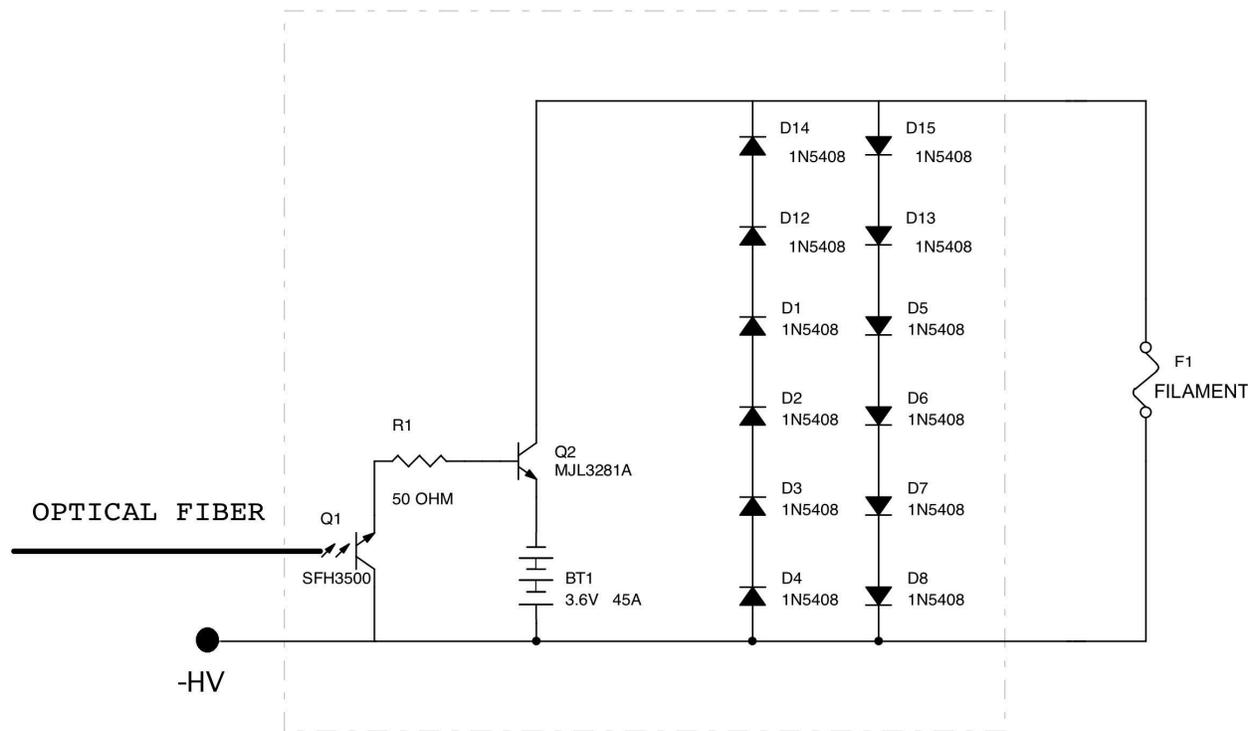}
 \caption{\small Electronic scheme of the floating filament heater circuit. \label{fig:FH}}%
 \end{figure}

A commercial HV power supply  (Spellman, $100\,$kV, $2\,$mA) is directly connected to the filament by means of HV feedthroughs ({\tt ft} in figure \ref{fig:blockdiagram}). 
The filament power supply is realized by using a pack of low-voltage batteries floated directly atop the HV source ({\tt bp\&fcc} in figure \ref{fig:blockdiagram}) and connected to the filament as shown in figure \ref{fig:FH}. The battery pack actually consists of three sets of fifteen $1.2\,$V rechargeable, Nickel-Metal-Hydride (NiMH) batteries connected  in parallel to each other whereas the three sets are connected in series so as to yield a total voltage drop of $3.6\,$V. The filament current is determined by transistors Q1 and Q2 in Darlington configuration. Q1 is a phototransistor that is optically driven by the light intensity of a superluminescent diode through an optic fiber ({\tt of} in figure \ref{fig:blockdiagram}). In this way, the HV floated battery pack is electrically decoupled from the low voltage part of the electron gun.

\subsection{General description and working principle}\label{sec:desc}
\begin{figure}[htbp]
 \includegraphics[width=\textwidth]{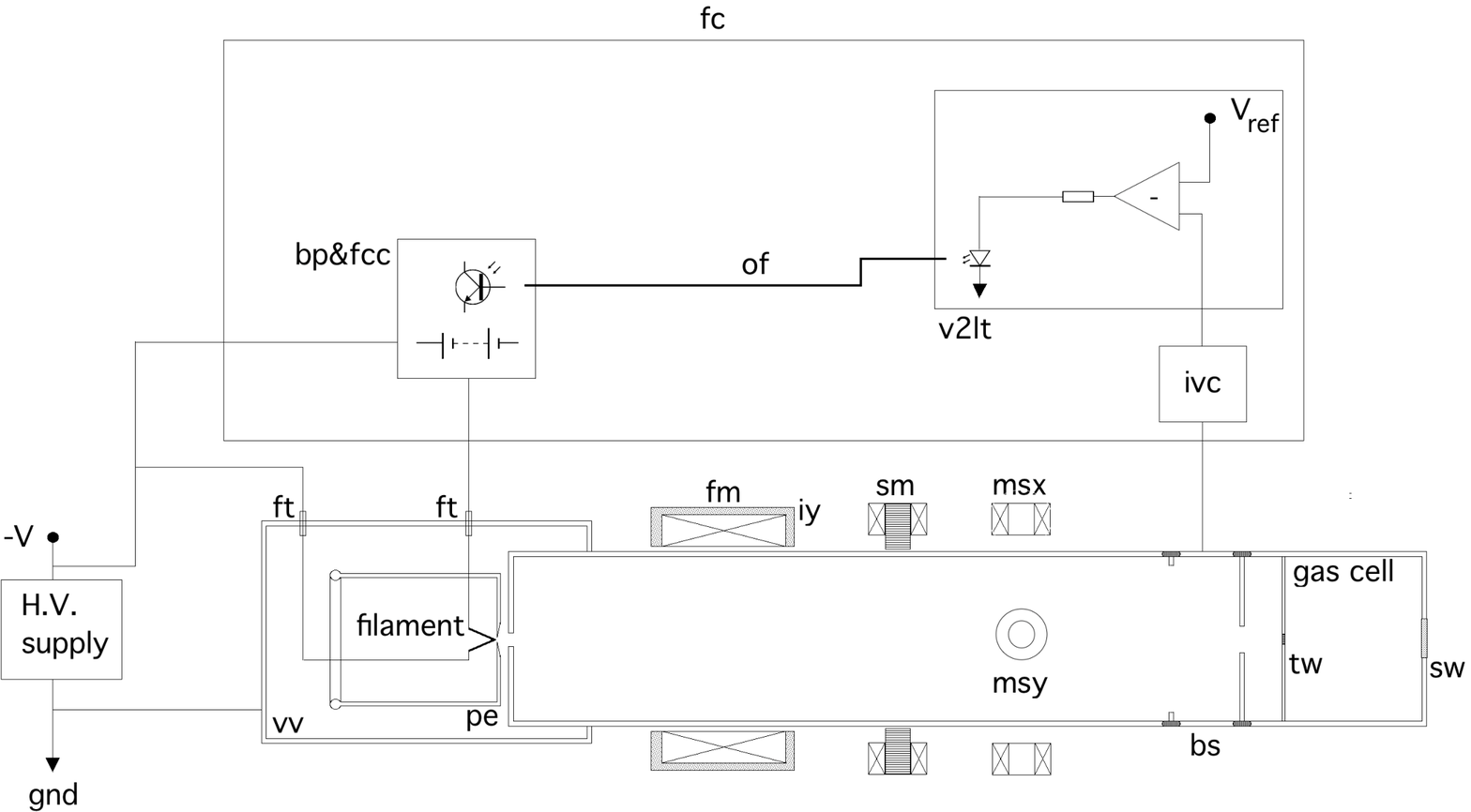}%
 \caption{\small Block diagram of the electron gun assembly. Legend: {\tt vv}= vacuum vessel, {\tt ft}=high-voltage feedtrough, {\tt pe}= Pierce electrode, {\tt fm}= focusing magnet, {\tt iy}= iron yoke, {\tt sm}= magnets for the sweeping of the electron beam, {\tt msx, msy} = magnetic stirrers for fine tuning of the $x$-$y$ position of the electron beam center, {\tt bs}= insulated section of the evacuated tube acting as beam stopper, {\tt tw}= Ti window, {\tt sw}= sapphire window, {\tt fc}= feedback control for the stabilization of the electron gun current, {\tt ivc}= current-to-voltage converter, {\tt V$_{\mathrm{ref}}$}= setting point of the electron beam current, {\tt v2lt}=voltage-to-LED light intensity converter, {\tt of}= optic fiber, {\tt bp\&fcc}= high-voltage floated battery pack and filament current control circuitry (not to scale).
 \label{fig:blockdiagram}}
 \end{figure}

The  filament is located in a vacuum vessel that is evacuated down to a limiting pressure of  $\approx 10^{-7}\,$mbar and is contained in a stainless steel cathode cup so that it is surrounded by an equipotential enclosure to prevent discharges. The filament tip faces a small opening in the cathode cup whose outer face ({\tt pe} in figure \ref{fig:blockdiagram}) is tapered to an angle of $\approx 20^{\circ}$ to focus electrons from the cathode to a point near the anode aperture so that nearly the whole electron beam is transmitted \cite{pierce1940}.
The anode plate is held at ground potential as the whole vacuum vessel and is is structurally mounted on the flange closing the end of the electron gun chamber. The edges of all metal surfaces exposed to high voltage are rounded so as to prevent discharges.

The electrons emitted from the cathode enter an evacuated tube through a circular aperture of $\approx 5\,$ mm diameter. X-rays are generated by fringe electrons impinging on the metal surfaces  so the whole electron gun is located inside of a Pb-shielded cabinet.

Upon passing through the anode aperture, electrons are injected into a $\approx 30\,$cm long, evacuated tube connecting the vacuum vessel to the cell. A first magnetic coil (labeled as {\tt fm} in figure \ref{fig:blockdiagram}) enclosed in an outer iron yoke ({\tt iy}) acts as a magnetic lens and provides focusing of the electron beam. 
The focalization provided by this coil depends on the electron energy and the dc current in the coil has to be manually adjusted when the accelerating high voltage is changed.

Our experiment on IR luminescence of rare gas excimers \cite{Borghesani2001,borghesani2007jpb} requires that the charge injection into the  gas is pulsed whereas the current emitted from the filament is continuous. In order to achieve a pulsed charge injection, we have introduced a magnetic coil wound around a core of grain-oriented steel laminations ({\tt sm} in figure \ref{fig:blockdiagram}) that is powered by a homemade power supply that can be operated at a frequency up to $ 1\,$kHz. The magnetic field generated by this coil harmonically deflects the electron beam around its axis so that charge is injected into the cell only when the beam crosses the cell entrance window. The duration of the charge pulse can be controlled by varying amplitude and frequency of the swing voltage across the coil windings. 

Two more pairs of crossed magnetic coils wound in air ({\tt  msx} and {\tt msy} in figure \ref{fig:blockdiagram}) are used to finely tune the position of the beam center with respect to the cell entrance window in order to make the shape of the charge pulse as symmetric as possible.

Before entering the sample cell, electrons cross an insulated section of the evacuated tube that acts as a Faraday cup and beam stopper ({\tt bs} in figure \ref{fig:blockdiagram}). This section of the tube is connected to a current-to-voltage converter ({\tt ivc} in figure \ref{fig:blockdiagram}). When electrons are deflected by the sweep coil and impinge on the beam stopper a voltage signal proportional to the electron current is generated by the current-to-voltage converter. This voltage signal is used as a feedback signal to stabilize the current intensity, as will be explained in the next section.

Electrons enter the sample cell through a circular window ({\tt tw} in figure \ref{fig:blockdiagram}) of $\approx 2\,$mm diameter made of a $\approx 7\,\mu$m thin, work-hardened Ti sheet. This window separates the high-vacuum section of the electron gun from the sample cell in which the gas pressure can be as high as $3\,$MPa. The thickness of the Ti window is small enough not to introduce too significant energy losses due to electron scattering inside the metal. However, electron scattering in the Ti foil spreads the injected charge  over a larger region, thus making the light emission more uniform.

Finally, the light emitted by the decaying excimers produced by electron impact finally exits the cell thorugh a sapphire window  ({\tt sw} in figure \ref{fig:blockdiagram}) in order to be analyzed by a Fourier-Transform IR interferometer \cite{Borghesani2001}.

\subsection{Electron gun current stabilization circuit}\label{sec:EGSC}
The life of the battery pack is approximately 12 hrs but its performance degrades over time leading to a slow but steady decrease of the electron beam current. This fact represents a major drawback for our experiment because the recording of a single IR spectrum may last up to 4 hrs. The IR signal stability and, hence, that of the charge per pulse injected into the gas must be quite high in order to obtain a reliable spectrum.

So, we have devised a feedback control that stabilizes the electron beam current by using the beam stopper signal as a feedback source. In the block diagram of figure \ref{fig:blockdiagram}, the feedback circuit block is labeled as ({\tt fc}). The current measured at the beamstopper is converted to a voltage by the block ({\tt ivc}) and is compared with a reference voltage ({\tt V$_{\mathrm {ref}}$}). This difference signal sets the working point of a superluminescent  diode whose light is coupled to the phototransistor Q1 in figure \ref{fig:FH} by means of the optical fiber. If the current collected at the beam stopper decreases with time, more light is emitted by the diode, thereby forcing an increase of the filament current supplied by the Darlington pair.

\begin{figure}[htbp]
 \includegraphics[width=\textwidth]{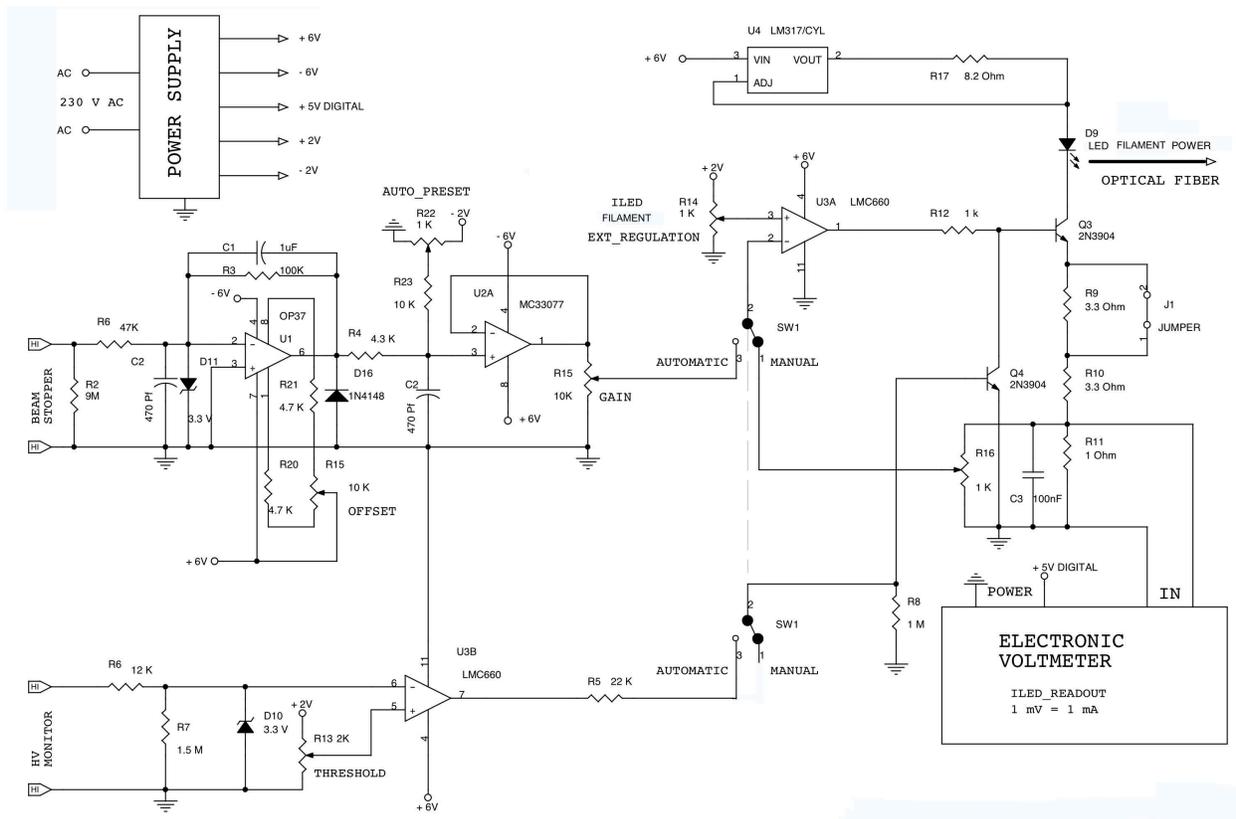}%
 \caption{\small Electronic scheme of the feedback circuit for the stabilization of the filament current. \label{fig:bpfcc}}%
 \end{figure}

In figure \ref{fig:bpfcc} a detailed electrical scheme of the feedback circuit is provided. The current collected at the beam stopper flows through the $4.7\,$k$\Omega$ resistor R6 and is time-averaged by the integrator consisting of op-amp OP37 and the resistor R3 and capacitor C1. When the switch SW1 is set to automatic, the beam stopper signal is subtracted from the reference voltage at op-amp U3A. This difference drives the transistor Q3 that sets the current in the superluminescent diode D9.
If switch SW1 is set to manual, no feedback is provided. 

In case of a discharge, as the beam stopper signal goes to zero, the feedback circuit would indefinitely keep increasing the light emitted by the diode. In order to prevent this, a voltage proportional to the high voltage, issued by the power supply itself, drives the comparator U3B. If the HV shuts down, the output of U3B goes high and turns transistor Q4 on, thus switching transistor Q3 off. In this way, feedback is suppressed and the filament is turned off.

\section{Electron gun performance}\label{sec:perf}

\begin{figure}[htbp]
 \includegraphics[width=\textwidth]{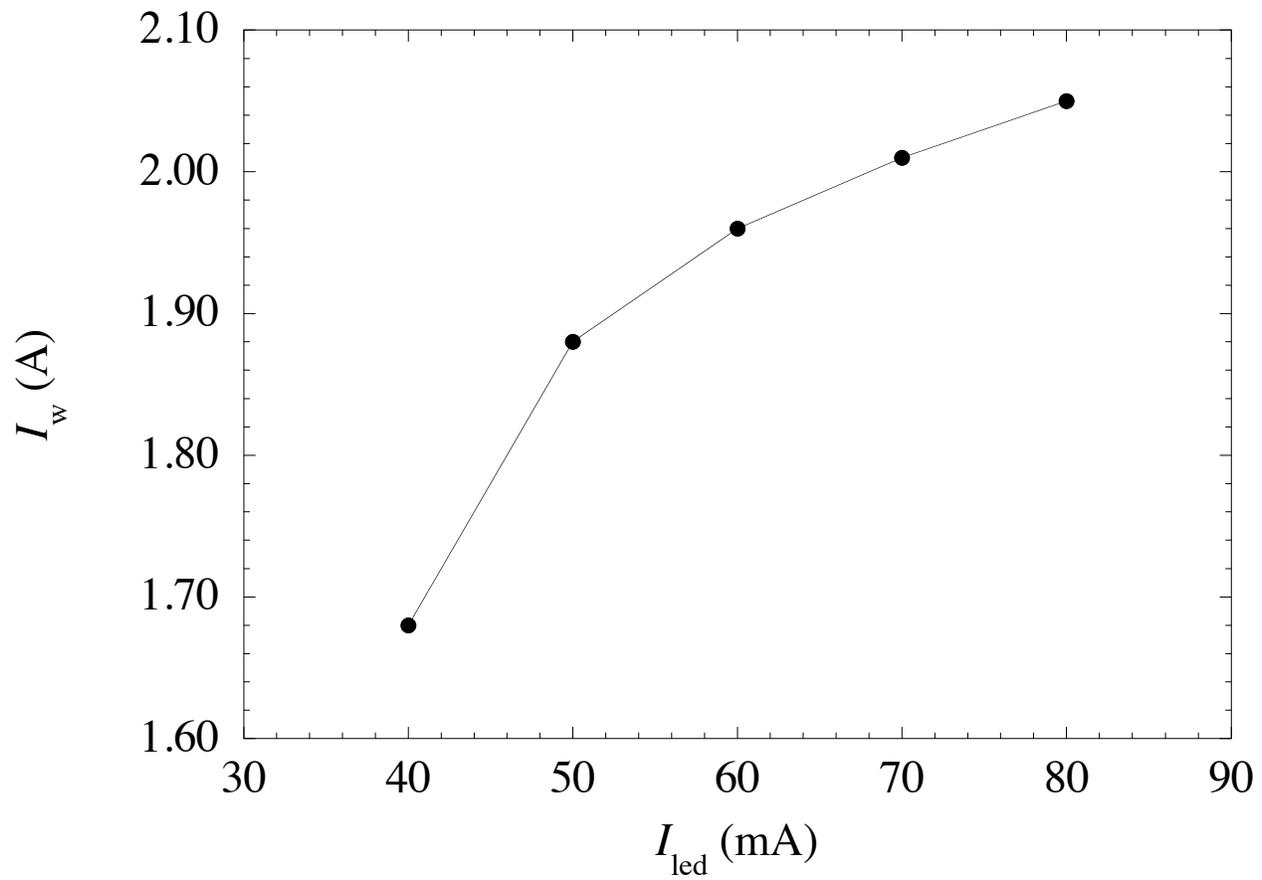}
 \caption{\small Filament current as a function of the LED current.  \label{fig:IiflIled}}
 \end{figure}

In this section we show the performance of the electron gun.
In figure \ref{fig:IiflIled} we report the filament current $I_{w}$ as a function of the current of the superluminescent diode. Current intensities in excess of $2\,$A are very easily obtained. For these current levels, the filament glows very brightly.

\begin{figure}[htbp]
 \includegraphics[width=\textwidth]{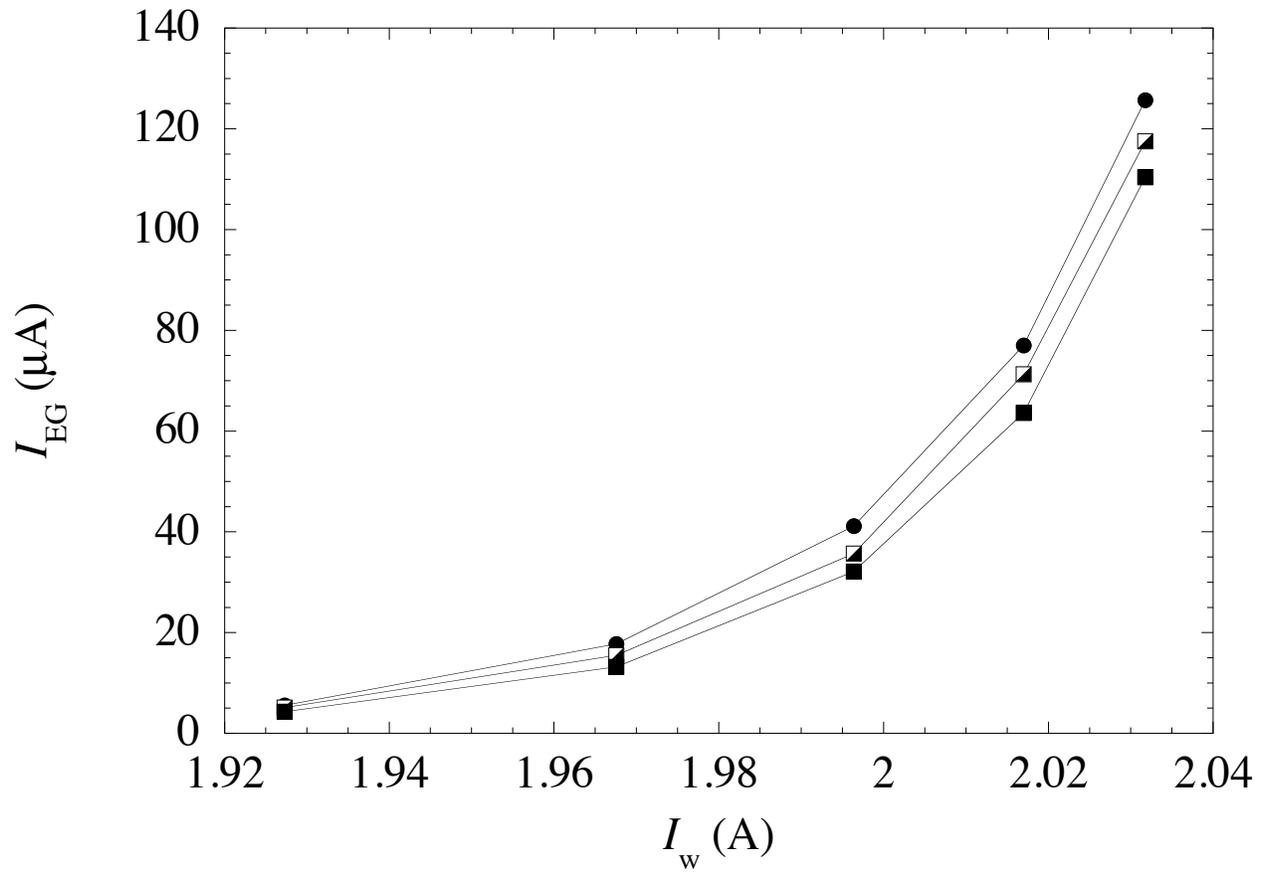}
 \caption{\small Current delivered by the electron gun as a function of the filament current for several values of the applied high voltage (from top: 70, 60, and 40 kV, respectively) at $f=100\,$Hz repetition rate.  \label{fig:IEGIw}}
 \end{figure}

In figure \ref{fig:IEGIw} we show the continuous current delivered by the electron gun, $I_\mathrm{EG},$ as a function of the filament current $I_{w}$ when the sweep frequency is set at $f=100\,$Hz for several values of the accelerating high voltage. The electron beam current increases rapidly with the current flowing through the filament and the extraction efficiency slightly increases with the accelerating high voltage.

\begin{figure}[htbp]
 \includegraphics[width=\textwidth]{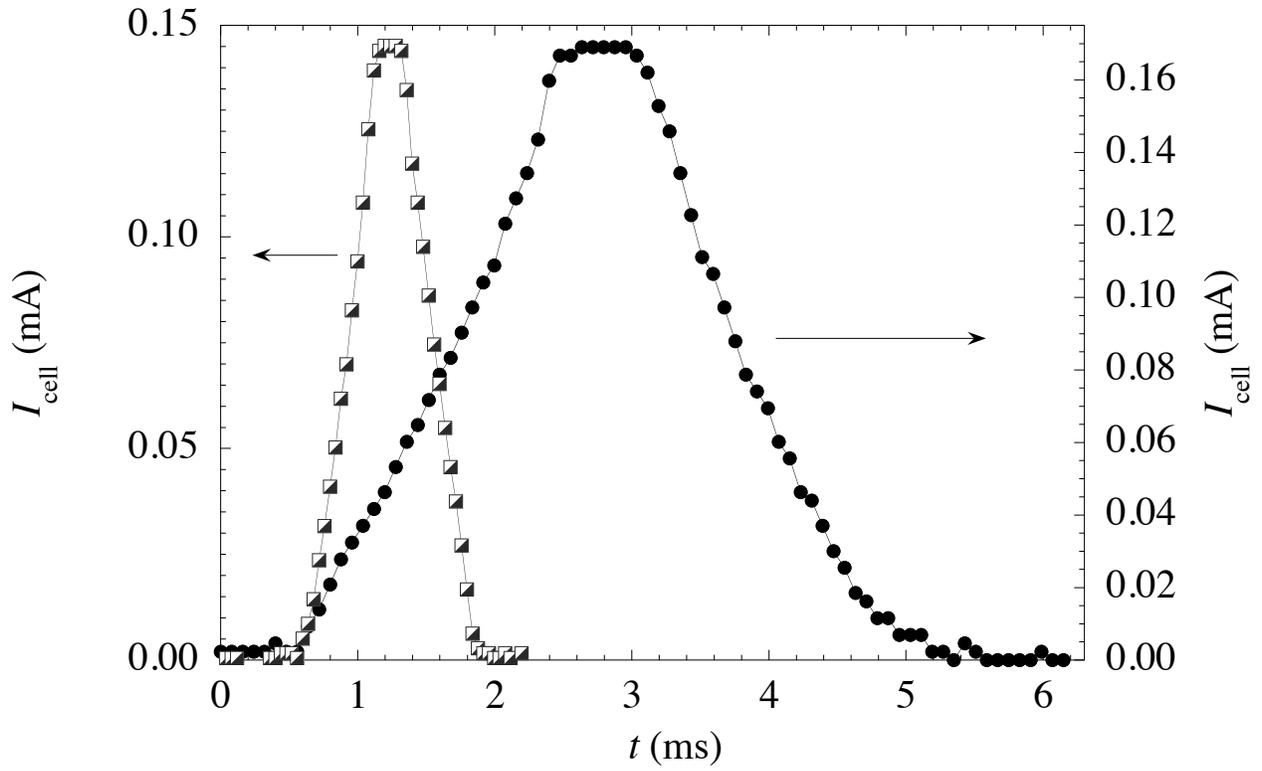}
 \caption{\small Shape of the current pulse injected into the cell at 100 (right scale) and 200 (left scale) Hz repetition rate for $ I_{w}=2.04\,$A at 70 kV.  \label{fig:pulseshape}}
 \end{figure}

In figure \ref{fig:pulseshape} we report the shape of the charge pulse injected into the cell for $I_{w}=2.04\,$A at 70 kV for $f=100$ and $200\,$Hz. The (quite symmetrical) trapezoidal shape of the pulse tells that the beam spot is smaller than the Ti window $(\approx 3\,$mm$^{2}).$ The steepness of the leading and trailing shoulder of the pulse can be controlled by either changing the sweep frequency or amplitude. 

If the sweep rate is $f,$ the beam spot passes $2f$ times per second across the Ti window. The two charge pulses obtained per each sweep cycle can be made symmetric in time and equal in amplitude by carefully adjusting the current intensity in the fine tuning magnetic coils {\tt msx} and  {\tt msy} in figure \ref{fig:blockdiagram}. By so doing, the interferometer can acquire two signals per sweep cycle, thus actually halving the time required to record a spectrum.

\begin{figure}[htbp]
 \includegraphics[width=\textwidth]{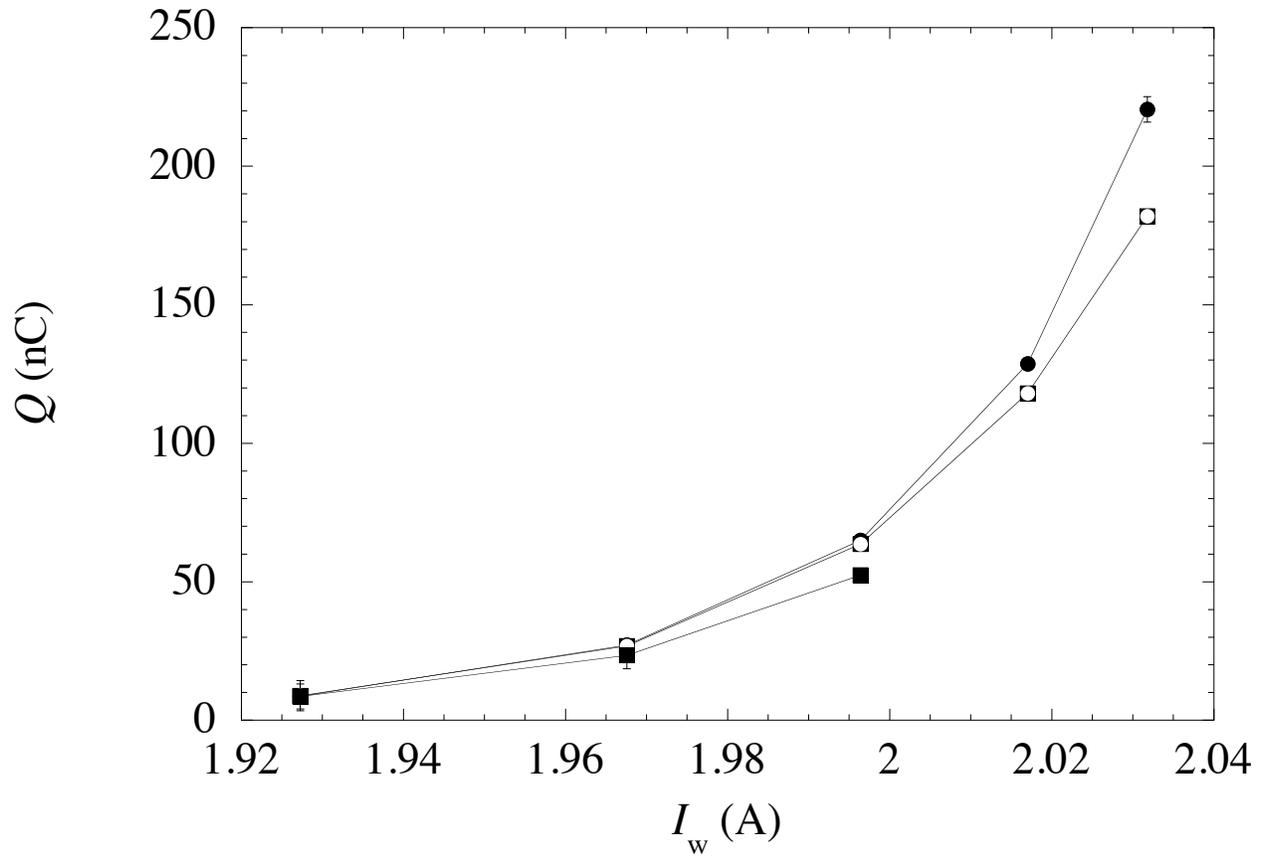}
 \caption{\small Charge per pulse injected into the cell as a function of the filament current for several values of the high voltage (from top: 70, 40, and 30 kV, respectively) at $f=100\,$Hz repetition rate.  \label{fig:Qcell}}
 \end{figure}

Another important quantity to be known is the amount of charge injected in the cell during each pulse. This quantity is proportional to the amount of luminescence light that can be produced. In figure \ref{fig:Qcell} we show the charge per pulse $Q$ injected into the sample cell as a function of the filament current $I_{w}$ at a sweep frequency $f=100\,$Hz for several high voltage values. It can be observed that several tens of nC can be easily injected into the cell, to be compared with the few tens of pC that were injected into the cell when using the previous electron gun.

Finally, we want to show how the feedback effectively stabilizes the performance of the electron gun. The electron gun stability has to be verified by checking the stability of the final outcome of the experiment, namely the final output of the interferometer. 

The interferometer is of the Michelson type \cite{born}. In the {\it stepscan} mode, the movable mirror is moved in discrete steps and the light signal is recorded at each step, yielding the so called {\it interferogram}. The spectrum is obtained by numerically performing the Fourier transform of the interferogram. So, the time stability of the interferogram has to be verified.

\begin{figure}[htpb]
 \includegraphics[width=\textwidth]{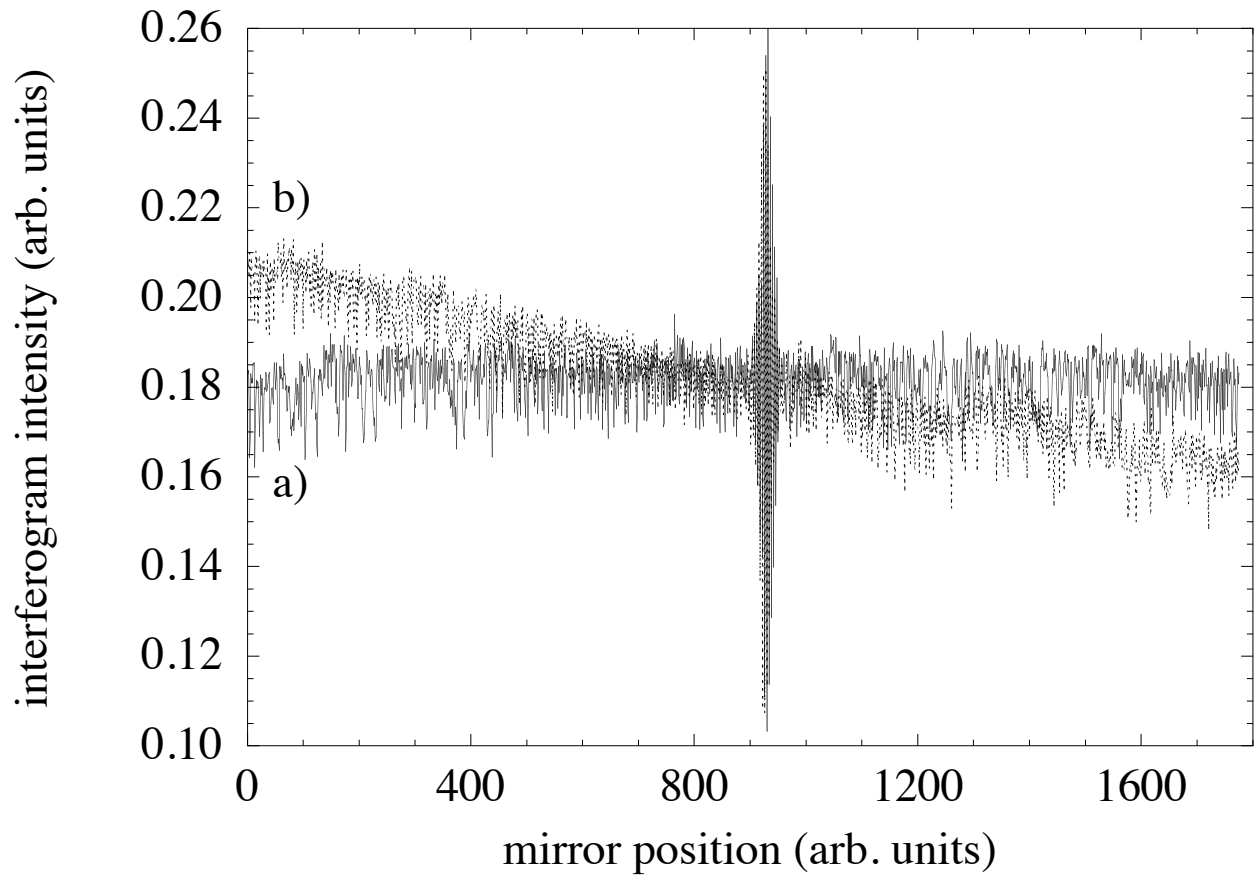}%
 \caption{\small Sample interferogram recorded with filament current stabilization (curve a) and without it (curve b).  The time interval over which the interferogram is recorded is approximately 3 hours.\label{fig:interferogram}}%
 \end{figure}

In figure \ref{fig:interferogram} we show two inteferograms. One is obtained without filament current stabilization (curve b), whereas curve a is obtained with filament current stabilization feedback. Approximately three hours were necessary to record each interferogram. The interferogram obtained without feedback clearly shows a decline of its amplitude with time. The fractional decrease is $\approx 8 \%/$hr. On the contrary, the interferogram obtained with feedback does not show any appreciable time variation. In any case, the feedback does not alter the position of the centerburst at which the maximum constructive interference takes place.

\section{Conclusions}\label{sec:conc}

We have devised a hot-filament electron gun to be used in our experiment of IR emission of rare gas excimers. Two are the main and novel features of this electron gun.
The first one is that both the HV  and the filament power supply are dc power supplies. The filament power supply is build by floating atop the HV a pack of low voltage batteries that can deliver large dc currents for several hours. The filament current is controlled by adopting an optical coupling scheme, so actually decoupling the high- and low-voltage sections of the electron gun.

The second most relevant feature is that the beam current intensity is monitored in order to realize a feedback that stabilizes the current itself which, otherwise, would slowly decrease in time owing to the batteries performance.

Our main goal, namely the strong increase of the amount of charge injected per pulse into the cell with respect to the previous electron gun used, has been successfully reached and the resulting device turns out to be reliable and easy and quite cheap to build. It is quite versatile because it can be used both in continuous as well as in pulsed mode only by exploiting a sweep magnetic coil.

At present, we are developing a system consisting of a triple-junction photovoltaic cell powered by a $1\,$W diode laser in order to replace the rechargeable batteries.
%



\newpage
 \clearpage
 \clearpage
 \clearpage
 \clearpage
\clearpage
\clearpage
 \clearpage
\clearpage

\end{document}